\documentclass{aipproc}

\layoutstyle{8d}

\usepackage[T1]{fontenc}
\usepackage[latin1]{inputenc}

\usepackage{amsmath}
\usepackage{graphicx}
\usepackage{natbib}

\begin{document}

\title{Different kinds of long-term variability from Cygnus~X-1}

\author{S.~ Benlloch}{
  address={IAAT -- Astronomie, Sand 1, 72076 T\"ubingen, Germany}
}
\author{K.~Pottschmidt}{
  address={MPE, Giessenbachstr. 1, 85748 Garching, Germany},
  altaddress={ISDC, Ch. d'\'Ecogia 16, 1290 Versoix, Switzerland}
}
\author{J.~Wilms}{
  address={IAAT -- Astronomie, Sand 1, 72076 T\"ubingen, Germany},
  altaddress={Department of Physics, University of Warwick, Coventry
  CV4 7AL, UK}
}
\author{M.A.~Nowak}{
  address={MIT/CSR, Cambridge, MA 02139, USA}
}
\author{T.~Gleissner}{
  address={IAAT -- Astronomie, Sand 1, 72076 T\"ubingen, Germany}
}
\author{G.G.~Pooley}{
  address={Cavendish Laboratory, Madingley Road, Cambridge CB3 0HE, UK}
}

\begin{abstract}
  We present a study of the long-term variability of Cyg X-1 using
  data from the RXTE/ASM and the RXTE/PCA during the time between the
  two soft states of 1996 and 2001/2002. This period has been
  characterized by many short ASM flaring episodes which we have
  identified as "failed state transitions". The 150\,d period which
  has been seen before and shortly after the 1996 soft state is not
  obviously present in the ASM rate during most of this time. Applying
  selection criteria from our pointed RXTE/PCA observations to exclude
  the flaring episodes we show that the 150\,d period can indeed still
  be significantly detected in the hard state. Furthermore, while the
  $\sim$420\,d timescale associated with the flaring is reduced in the
  selected hard state count rate, it is still pronounced in the
  temporal evolution of the corresponding hardness ratios. The Ryle
  radio flux is also consistent with the 150\,d period being present
  but distorted during this time.
\end{abstract}

\maketitle

\section{Introduction \&  selection criteria}

\begin{figure*}
\centering
\includegraphics[width=1.45\columnwidth]{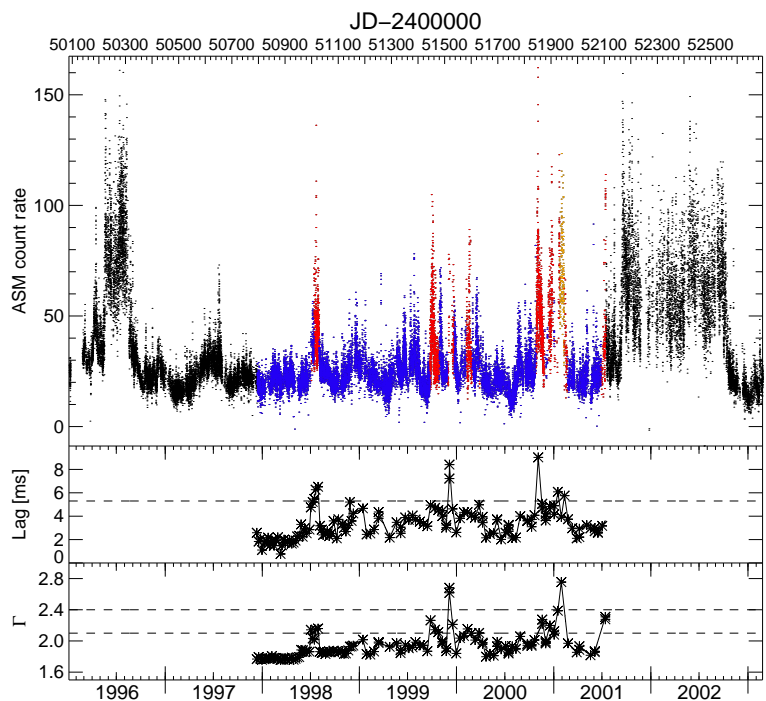}
\caption{Upper panel: RXTE/ASM 1.3--12.2\,keV lightcurve of
  Cyg~X-1. A categorization of the ASM data points into different
  states has been performed using the parameters and selection
  criteria displayed in the two lower panels: the parameters have been
  derived from regular monitoring observations of Cyg~X-1 with the
  RXTE/PCA and consist of the Fourier time lag between the 2--4\,keV
  and 8--13\,keV high resolution lightcurves (middle panel) as well as
  of the spectral photon index $\Gamma$ (lower panel). See text for an
  explanation of which selection criteria (horizontal lines) apply for
  which state.}\label{fig:asm}
\end{figure*}

Historically several different scales of long-term variability in
Cyg~X-1 have been reported. Multiwavelength observations (optical,
radio, hard and soft X-ray) between 1996 and 1998 established that a
150\,d signature is the dominant variability timescale which is
suggested to be caused by precession and/or radiative warping of the
accretion disk \citep{pooley:98a,brocksopp:99a}. Recently this period
has also been confirmed with earlier Ginga data \citep{kitamoto:00a}.
From 1998 onwards, however, this signature is not obvious in the soft
X-ray flux anymore (Fig.~\ref{fig:asm}, upper panel): Although mainly
in the hard state, Cyg~X-1 has been more active since before, i.e.,
more flaring episodes during which the source can be found in the
intermediate state (=failed state transitions \citep{pottschmidt:02a})
-- sometimes even reaching the soft state -- have been observed. This
goes together with a sharp and significant change in the properties of
the short-term variability in June 1998 \citep{pottschmidt:02a}.
\citet{oezdemir:01a} suggested to avoid the distortion of the possible
periodicities in the original data introduced by the high/soft state
and the failed state transitions by truncating the RXTE/ASM lightcurve
at 30\,cps (threshold derived from visual inspection). They report a
clearer detection of the 150\,d period in the 1996 to 1999 ASM data.
Based on the principal idea of selecting only hard state data, we:

\begin{itemize}
\item perform a period search for RXTE/ASM data dominated by the
  flare-distorted phase from 1998-2001,
\item use physical selection criteria derived from monitoring
  \mbox{Cyg~X-1} with RXTE/PCA since 1998,
\item use two statistics for unevenly sampled data (Lomb-Scargle PSD
  and epoch folding),
\item also study the variability of the RXTE/ASM colours. 
\end{itemize}

Our selection criteria (lower panels of Fig.~\ref{fig:asm}) are based
on the photon index $\Gamma$ of the PCA spectra and the time lags
(between the 2--4 and 8--13\,keV flux), since they have been shown to
be good diagnostics for failed state transitions
\citep{pottschmidt:00a,zdz:02a}. $\Gamma$ and lag are interpolated
from the PCA sampling to the ASM sampling. In detail the states are
defined as follows: (i) the canonical hard state is characterized by
hard PCA spectra with $\Gamma < 2.1$, (ii) spectra with $2.1 \leq
\Gamma \leq 2.4$ are typical for failed state transitions, and (iii)
for the softest spectra with $\Gamma > 2.4$, enhanced time lags $\geq
5.3$\,ms also point towards a failed state transition, while
``normal'' time lags $< 5.3$\,ms characterize the soft state.

\section{Long-term variability, significances}

\begin{figure}
\resizebox{\columnwidth}{!}
{\includegraphics{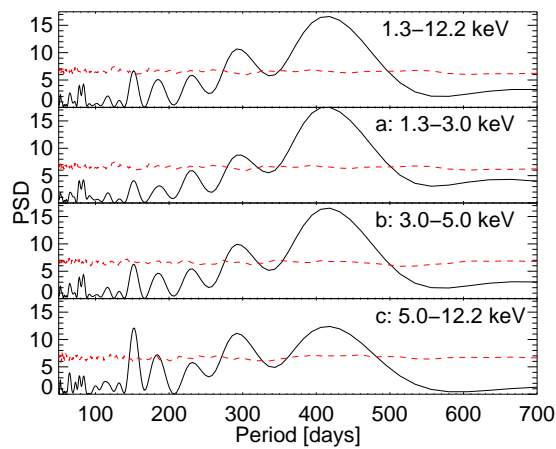}} 
\caption{Lomb-Scargle periodogram of all Cyg X-1 ASM data from 1997
  December to 2001 July. No state selection criteria have been
  applied. The first panel shows the result for the full ASM energy
  range while the other three panels are derived for the indicated
  sub-bands. }\label{fig:all}
\end{figure}

Fig.~\ref{fig:all} shows the Lomb-Scargle periodogram
\citep{scargle:82,lomb:76a} for all ASM data between 1997 December and
2001 July. The analysis of the total count rate as well as of the
three ASM sub-bands is displayed. Prior to the analysis, the
lightcurves have been rebinned to a 7\,d resolution in order to avoid
aliases due to the 5.6\,d orbital period. The dashed lines represent
the $99.9\%$ ``local significance'' levels for a set of 5000 Monte
Carlo white noise simulations (for more details on the significance
determination see \citep{benlloch:01b}). As expected, the 150\,d
signature is not significantly present in the overall ASM data.
Longer timescales seem to be more prominent especially a broad peak
around ~420\,d (of course, increasingly fewer cycles are covered).
However, there is already an indication here that the 150\,d period is
still present from the data in the 5-12\, keV energy band, also
strengthening the idea of possible inner disk / radio jet coupling.

\section{Is the 150\,d scale still there?}

\begin{figure}
\resizebox{\columnwidth}{!}
{\includegraphics{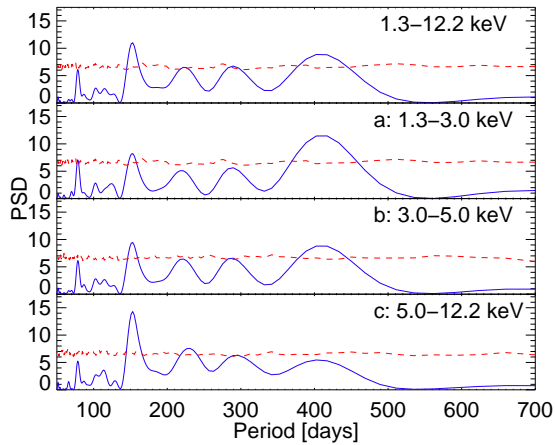}} 
\caption{The same as Fig.~\ref{fig:all} but for the hard state
  selected ASM data only.}\label{fig:scargle}
\end{figure}

\begin{figure}
\resizebox{\columnwidth}{!}
{\includegraphics{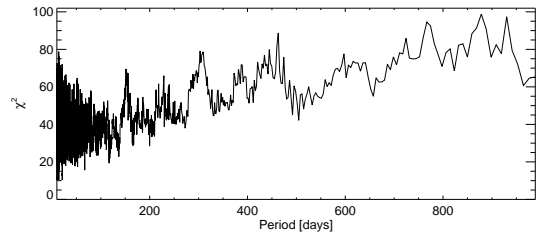}} 
\caption{Results of a period search for the hard state selected ASM
  data using the epoch folding method.}\label{fig:epfold}
\end{figure}

After applying the selection criteria described above to keep only the
hard state data, two types of changes are apparent from the resulting
Lomb-Scargle PSDs (Fig.\ref{fig:scargle}): the significance of the
150\,d period increases in all energy bands and it can now be
significantly detected in the total count rate. In parallel the
opposite behavior is displayed by the 420\,d component: compared to
the unselected dataset its significance decreases in all energy bands,
most strongly in the hardest band where it decreases below the given
significance level. These two trends lead to the 150\,d signature
being the most prominent timescale of the total ASM rate in the
cleaned hard state. Its presence is also confirmed by the results of
an epoch folding period search, see Fig.\ref{fig:epfold}.

\section{Confirmation from the radio flux}

\begin{figure}
\includegraphics[width=\columnwidth]{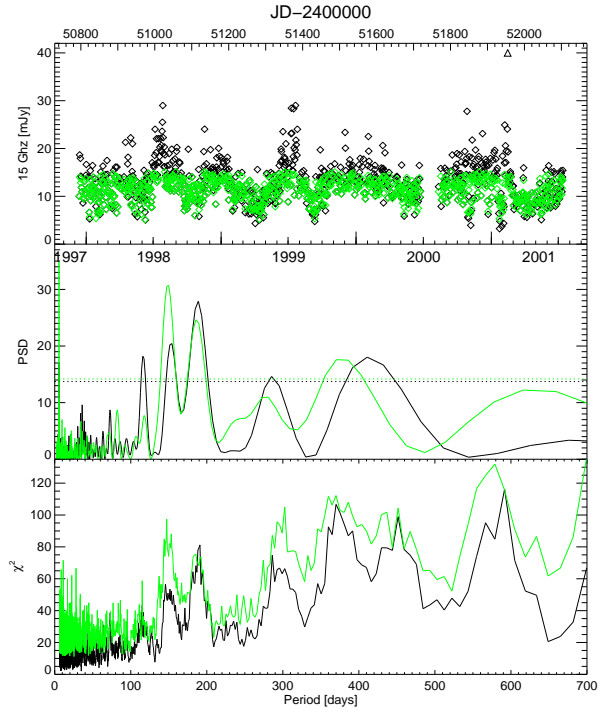}
\caption{Upper panel: 15\,GHz radio lightcurve measured by the Ryle
  telescope. The full data set (dark symbols) as well as the hard
  state selected one (light symbols) is shown. See text for details on
  the flux selection criterion applied in this case. Middle panel:
  Lomb-Scargle PSDs for the Ryle lightcurves. Here the $99.9\%$
  ``global significance'' levels are shown since they are not expected
  to deviate from the local significances. Bottom panel: Epoch folding
  analyses for the Ryle lightcurves.}\label{fig:radio}
\end{figure}

While the analysis of the overall 15\,GHz radio lightcurve --
performed over the same time interval as the ASM analysis -- does show
the presence of the 150\,d signature as a formally significant
timescale, other nearby timescales arise with comparable significance
(Fig.~\ref{fig:radio}, dark symbols and lines). Since we are studying
a non-canonical hard state (changed short-term X-ray variability
\citep{pottschmidt:02a}) it is not a priori clear that a selection
based on the X-ray states is also appropriate for the radio emission
-- although the tendency for radio flaring during failed state
transitions is known. In this first approach we thus use a flux
selection criterion (5\,mJy $<$ Ryle flux $<$ 15\,mJy) and find that
it helps to better work out the 150\,d timescale
(Fig.~\ref{fig:radio}, light symbols and lines). The overall picture
of the 150\,d period being present but buried by the flaring activity
is thus consistent. However, from this radio dataset alone and using
the preliminary flux selection criterion one could not determine the
physical reality of the 150\,d period as compared to, e.g., the 190\,d
variability. Note that a 1d-binned-lightcurve was analyzed here --
leading to the presence of the 5.6\,d orbital period in the PSD -- and
that the described effects are also present using 7d-bins, but are
less pronounced.

\section{A different spectral timescale?}

\begin{figure}
\includegraphics[width=0.9\columnwidth]{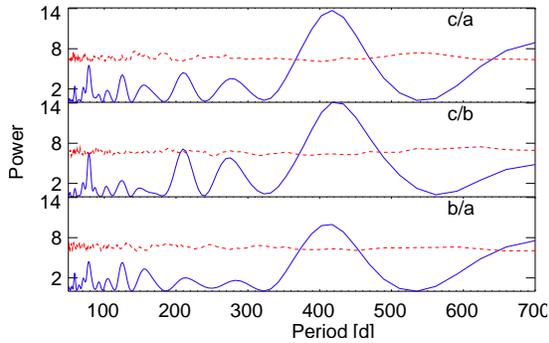}
\caption{Lomb-Scargle periodogram of the colours derived from the hard
  state selected ASM data. The energy bands a, b, c used for the
  colour ratios are defined in
  Fig.~\ref{fig:all}).}\label{fig:colours}
\end{figure}

It is also interesting to perform the same period search on the ASM
colours, applying the hard state selection criteria
(Fig.~\ref{fig:colours}): the 150\,d period is not visible at all,
while the 420\,d timescale is clearly dominating. We cautiously
suggest that while the 150\,d period is a pure intensity variation,
there might be another characteristic timescale driving the soft X-ray
flux, associated with spectral changes. Note that this is true even
after removing the failed state transitions themselves.

\begin{theacknowledgments}
  This work has been financed by DLR grant 50~OX~0002 (SB) and by DFG
  grant Sta~173/25-3 (KP, TG). KP acknowledges support from Chandra
  grant GO-4050B and from the Soci\'et\'e Suisse d'Astrophysique et
  d'Astronomie for enabling her to attend the ``X-Ray Timing 2003:
  Rossi and Beyond'' conference in order to present this work. The
  co-authors dedicate the presentation of this work to the twins, Eric
  and Marvin, who timed their birth allowing their mother (SB) to
  finish the analysis but suggesting that its presentation should be
  done by one of the collaborators (KP).

\end{theacknowledgments}

\bibliographystyle{aipproc}

\begin{thebibliography}{10}
\expandafter\ifx\csname natexlab\endcsname\relax\def\natexlab#1{#1}\fi
\providecommand{\enquote}[1]{``#1''}
\expandafter\ifx\csname url\endcsname\relax
  \def\url#1{\texttt{#1}}\fi
\expandafter\ifx\csname urlprefix\endcsname\relax\def\urlprefix{URL }\fi

\bibitem[Pooley et~al.(1999)]{pooley:98a}
Pooley, G.~G., Fender, R.~P., and Brocksopp, C., \emph{MNRAS}, \textbf{302}, L1
  (1999).

\bibitem[Brocksopp et~al.(1999)]{brocksopp:99a}
Brocksopp, C., Fender, R.~P., Larianiov, V., Lyuty, V.~M., Tarasov, A.~E.,
  Pooley, G.~G., Paciesas, W.~S., and Roche, P., \emph{MNRAS}, \textbf{309},
  1063 (1999).

\bibitem[Kitamoto et~al.(2000)]{kitamoto:00a}
Kitamoto, S., Egoshi, W., Miyamoto, S., Tsunemi, H., Ling, J.~C., Wheaton,
  W.~A., and Paul, B., \emph{ApJ}, \textbf{531}, 546--552 (2000).

\bibitem[Pottschmidt et~al.(2003)]{pottschmidt:02a}
Pottschmidt, K., Wilms, J., Nowak, M.~A., Pooley, G.~G., Gleissner, T., Heindl,
  W.~A., Smith, D.~M., Remillard, R., and Staubert, R., \emph{A\&A},
  \textbf{407}, 1039--1058 (2003).

\bibitem[Oezdemir and Demircan(2001)]{oezdemir:01a}
Oezdemir, S., and Demircan, O., \emph{Ap\&SS}, \textbf{278}, 319--325 (2001).

\bibitem[Pottschmidt et~al.(2000)]{pottschmidt:00a}
Pottschmidt, K., Wilms, J., Nowak, M.~A., Heindl, W.~A., Smith, D.~M., and
  Staubert, R., \emph{A\&A}, \textbf{357}, L17--L20 (2000).

\bibitem[Zdziarski et~al.(2002)]{zdz:02a}
Zdziarski, A.~A., Poutanen, J., Paciesas, W.~S., and Wen, L., \emph{ApJ},
  \textbf{578}, 357--373 (2002).

\bibitem[Scargle(1982)]{scargle:82}
Scargle, J.~D., \emph{ApJ}, \textbf{263}, 835 (1982).

\bibitem[Lomb(1976)]{lomb:76a}
Lomb, N.~R., \emph{Ap\&SS}, \textbf{39}, 447--462 (1976).

\bibitem[Benlloch et~al.(2001)]{benlloch:01b}
Benlloch, S., Wilms, J., Edelson, R., Tahir, Y., and Staubert, R., \emph{ApJ},
  \textbf{562}, L121--L124 (2001).

\end{thebibliography}

\end{document}